\input epsf.tex
\documentstyle [12pt] {article}

\begin{document}

\baselineskip 7.5mm

\begin{flushright}
\begin{tabular}{l}
IOA-296    \\
gr-qc/9401023 \\
Oct. 1993  
\end{tabular}
\end{flushright}
\vspace{4mm}
\begin{center}
{\Large \bf On the existence of singularity-free solutions  }\\
\vspace{2mm}
{\Large \bf  in quadratic
gravity } \\
\vspace{5mm}

J.Rizos\footnote{Research supported by EEC contract
Ref.B/SCI*-915053}\\
Centre de
Physique Theorique
\footnote{Laboratoire Propre du Centre National de la Recherche
Scientifique UPR A.0014.}
, Ecole Polytechnique, 91128, Palaiseau, FRANCE\\

and\\
 K. Tamvakis\\
Physics Department, University of Ioannina , GR-45110 Ioannina,
GREECE\\
\vspace{6mm}

\vspace{20mm}

{\bf Abstract}
\end{center}

We study a general field theory of a scalar field coupled to
gravitation through a quadratic Gauss-Bonnet term $\xi({\phi})$$R_{GB}^2$.
We show that, under mild assumptions about the function $\xi(\phi)$, the
classical solutions in a spatially flat FRW background include 
singularity - free solutions.

\vspace{50mm}

\newpage
\def\dom{\dot\omega}
\def\dphi{\dot\phi}

Recently it was found that the loop corrected superstring effective
action in a FRW background in the presence of the dilaton and modulus fields
admits a particularly interesting class of singularity-free solutions with
flat initial asymptotics$^{[1]}$.Singularity-free solutions have been
considered by other authors as well$^{[2],[3]}$ and their existence,
although linked to the presence of quadratic curvature terms, depends on
the form of the gravitational coupling to scalar fields.
 The  purpose of this
letter is to study the existence of singularity- free solutions and
formulate the necessary conditions that should be imposed on the
quadratic gravitational coupling.  Although motivated by the concrete
case of the superstring the analysis at hand is quite general and is
carried out in an effective field theory framework that allows a
possible alternative, at present non-existing, quantum gravitational
underlying theory. 
\par Consider a scalar field coupled quadratically to  gravity through
the Langrange density $$L = {1\over2} R + {1\over2} (D\phi)^2 -
{\delta\over16} \xi(\phi) R_{GB}^2\eqno(1)$$
where $R_{GB}^2=(R_{\mu\nu\kappa\lambda})^2-4(R_{\mu\nu})^2+R^2$ is the
standard Gauss-Bonnet integrand.The function $\xi(\phi)$ is a general 
smooth
function of the scalar field which will be further constrained
shortly. $\delta$ is just a coupling parameter. 
Note that we have chosen a
Gauss-Bonnet quadratic coupling in accordance with general
 unitarity arguments
as well as the concrete superstring case$^{[4]}$.
 In a standard RWF background
metric $g_{\mu\nu}=(1,-e^{2\omega}\delta_{ij})$ we obtain the classical
equations of motion
$$\phi'' +3\dom\dphi+{3\over2}\delta\xi'(\phi)
\dom^2(\omega''+\dom^2)=0\eqno(2)$$
$$3\dom^2-{1\over2}\dphi^2-{3\over2}\delta\xi'
(\phi)\dphi\dom^3=0\eqno(3)$$
In terms of $ x=\dphi$ and $z=\dom$ considered as functions of $\phi$,
equation (3) can be solved as an algebraic equation and give
$$x=z[-{3\over2}\delta\xi'z^2\pm\sqrt{6+({3\over2}
\delta\xi^{'}z^2)^2}\space]\eqno(4)$$
while (2) becomes$$z'\equiv{dz\over d\phi}=-{z^2\over x}(A/B)\eqno(5)$$
where
$$A\equiv16z^4+16z^2x^2+{20\over9}x^4-{8\over3}\delta\xi^{''}
z^2x^4\eqno(6)$$
$$B\equiv16z^4-{16\over3}z^2x^2+{20\over9}x^4\eqno(7)$$   
\par Let us assume now
that $\xi(\phi)$ is a smooth function that possesses  one minimum at
some point $\phi_0$ . For example $\xi$ could be $\phi^2$,$\phi^4$,... . 

We shall show that under the assumption that $\xi(\phi)$ grows 
 faster than $\phi^2$
at $\phi\rightarrow\pm\infty$ the system of equations of motion $(2),(3)$,
 admits
non-singular solutions for the case $\delta>0$. Let us outline here
the basic steps of our proof. We first show that all singularities occur at
$|z|\rightarrow{\infty}$. Expanding asymptotically our
 equations at this region and
analyzing the various possibilities we show that only one
 singular solution exists
$\omega\sim\log t, \phi\sim\phi_s+t^2/(\delta \xi'(\phi_s))$. Then
 we show that  
the points $z=0$ or $x=0$
 cannot be continuously approached for finite $\phi$ and
thus every solution is characterized by a definite sign of $z$ and $x$.
 Choosing
$z>0$ (expanding universe) and $x>0$ we show that
 the singularity point $\phi_s$ is
related to the $\xi$ minimum $\phi_0$, and we always have
  $\phi_s >\phi_0$, for
$\delta>0$. Furthermore,
 we show that a singular solution involves only points
$\phi>\phi_s$
 and
thus  any solution that includes a point $\phi<\phi_0$ is non-singular.
 Since such
a point can be always found this is an existence proof of the singularity free
solutions. Finally, in order to make the results of our analysis concrete,
we present the  numerical solutions of the case $\xi=\phi^2/2$, $\delta=1$.
 
 A ``singular solution" is one for
which $|z|\rightarrow{\infty}$ or $|\dphi|\rightarrow{\infty}$
  at some finite time.
Note that we could have  $z\rightarrow{\infty}$ for
 $\phi\rightarrow{\infty}$  and
still have a singularity at a finite time. We shall restrict the coupling
function $\xi(\phi)$ to increase at least as fast as $\phi^2$ at
infinity. Then, we can show that $x(\phi)$ cannot have a singularity
while $z$ is maintained at a finite value. In order to prove it we first
solve equation (4) for large $x$ and obtain $x\sim-3\delta\xi^{'}z^3$ .
Obviously $x$ has a singularity while $z$ is finite, only
if $\delta\xi^{'}(\phi)$ has one. Since $\xi(\phi)$ is supposed to be a
smooth function this can occur only at $\phi\rightarrow{\pm\infty}$ .
Next, we go to equation (5) which takes the
form
$$
z^{'}\sim({1\over3\delta\xi^{'}z})(1-{6\over5}\delta\xi^{''}z^2)
\eqno(8)
$$ 
 We can proceed considering three separate cases. Namely
 $\delta\xi^{''}z^2\gg$O(1) , $\delta\xi^{''}z^2\sim$O(1)  and
 $\delta\xi^{''}z^2\ll$O(1)  while always $\delta\xi^{'}z^2\gg$O(1) .
Solving (8) in the first case leads to the
contradiction $x\sim(\delta\xi^{'})^{-{1\over5}}\rightarrow0$. The second
case can be treated by setting $\delta\xi^{''}z^2=b$ and getting from
(8) the
solution
$z\sim\phi^{{1\over2}(1-\kappa)}$,
$x\sim\phi^{{3\over2}-{\kappa\over2}}$, 
$\delta\xi\sim\phi^{1+\kappa}$  for $\phi\rightarrow{\pm\infty}$
 with $\kappa\equiv({1\over5}+{2\over3b})^{-1}$. In order to have the
assumed behavior of  $z$,$x$ and $\delta\xi$ we have to constrain
$1\leq\kappa\leq3$. Solving in terms of time we obtain
$\phi\sim{(t+...)}^{2\over(\kappa-1)}$  . Thus, the point
$\phi\rightarrow\pm\infty$ is not approached at a finite time and we do
not have a true singularity in this case. Finally, in the last case
$\delta\xi^{''}z^2\ll$O(1)  the equation (8) becomes
$z^{'}\sim{1\over{3\delta\xi^{'}z}}$  which is satisfied by
$z^2\delta\xi^{'}\sim{2\over3}\phi$  for $\phi\rightarrow\pm\infty$ .
Expressing (8) solely in terms of $z$ gives
$z^2\sim\phi$, $\delta\xi^{'}\sim$const. which contradicts  our
assumptions.

Let us now proceed by deriving the asymptotic form of
the singular solutions. We can start from equation (4) and solve it around the
singular point $\phi_s$ for which $z(\phi_s)\rightarrow\infty$ .
 Three different
possibilities arise in principle for $z$ near the
singularity, namely $\delta\xi^{'}z^2\sim$O(1),
$\delta\xi^{'}z^2\ll$O(1) and $\delta\xi^{'}z^2\gg$O(1) . They can be
analyzed as follows:  
       
1) $\delta\xi^{'}z^2\sim$O(1). In this case we set
${3\over2}\delta\xi^{'}z^2=a$  near $\phi_s$
 and obtain from (4) $x={\lambda}z$ with $a={3\over\lambda}-{\lambda\over2}$.
Equation (5) gives 
$${{z^{'}}\over{z}}={-{(1+\lambda^2+{{5\lambda^4}\over36}-
{{\lambda^4\delta\xi^{''}z^2}\over6})}
\over{{\lambda}(1-{{\lambda^2}\over3}+{{5\lambda^4}\over36})}}\eqno(9)$$
or ${{z^{'}}\over{z}}=-{{A}\over{{\lambda}B}}$ . For
$\delta\xi^{''}z^2=b\sim$O(1) we get
$z\sim{\exp(-{A\phi\over{{\lambda}B}})}$ which is singular at
$\phi\rightarrow\pm\infty$  and leads to
$\delta\xi^{'}\sim b \exp(-{2A\phi\over{{\lambda}B}})$  in contradiction
to our assumptions for the behavior of $\xi(\phi)$ at infinity. For
$\delta\xi{''}z^2\gg$O(1)  we get the solution
$z^{-2}\sim{{{{\lambda}^3}\delta\xi{'}\over{3B}}+c}$ 
 which requires c=0 and
${{\lambda}\over2}-{3\over{\lambda}}
={{9(1-{{{\lambda}^2}\over3}+{{5{\lambda}^4}\over36})}/2{\lambda^3}}$.
The last relation is impossible since there are no real solutions for $\lambda$.
Finally, the case $\delta\xi{''}z^2\ll$O(1) is just like the first case.    
  
 2) $\delta\xi{'}z^2\ll$O(1). In this case from equation (4) we obtain
$x\sim\pm{z}\sqrt{6}$ . Equation (5) becomes
$${z^{'}\over z}\sim\mp3(1-{\delta\xi^{''}z^2\over2})/\sqrt{6}\eqno(10)$$
 If $\delta\xi^{''}z^2=b\sim$O(1) we obtain
$z\sim\exp(\mp\sqrt{3\over2}(1-{b\over2})\phi)$  which
corresponds to a singularity only for $\phi\rightarrow\pm\infty$ . This
implies
$\delta\xi^{''}\sim\exp(\pm{2\sqrt{3\over2}}({1-{b\over2}})\phi)$ 
which is not in agreement with our assumptions for $\xi(\pm\infty)$.
 Similarly for
the case $\delta\xi^{''}z^2\ll$O(1). Finally,
 if $\delta\xi^{''}z^2\gg$O(1) , (10)
gives $z^{-2}\sim{\pm{\sqrt{3\over2}}{\delta\xi^{'}}}$  which is
incompatible with $\delta\xi^{'}z^2\ll$O(1).      
 
 3)$\delta\xi^{'}z^2\gg$O(1). In this case we get
two different solutions for $x$, namely $x_-\sim{-3\delta\xi^{'}z^3}$ 
and $x_+\sim{2\over{\delta\xi^{'}z}}$ . \par For $x=x_-$,
neccessarilly $\vert{x}\vert\gg\vert{z}\vert$  and equation (5)
becomes
$${z^{'}}\sim{({1-{{6\delta\xi^{''}z^2}\over5}})\over{3\delta\xi^{'}z}}
\eqno(11)$$ 
which is equivalent to
${({3\over2}\delta\xi^{'}z^2)^{'}}\sim{1+{{3\over10}\delta\xi^{''}z^2}}$ 
. If $\delta\xi^{''}z^2=b\sim$O(1) we obtain
${3\over2}\delta\xi^{'}z^2\sim{{(1+{3b\over10})\phi}+...}$ which
implies that the singularity appears for infinite $\phi$ . Then, we
get $\delta\xi^{'}\sim{\phi^{3b/{({2+{3b\over5}})}}}$ and
$z^2\sim{\phi^{({1-{6b\over5}})/({1+{3b\over10}})}}$,
 with $b<{5\over6}$ in order
to have $|z|\rightarrow\infty$ .
 The constraint $b<{5\over6}$ leads to a function
$\delta\xi$ that does not increase at infinity as fast as $\phi^2$, in
contradiction to our assumptions. If $\delta\xi^{''}z^2\ll$O(1), we obtain
${3\over2}\delta\xi^{'}z^2\sim\phi$  which also requires
$\phi\rightarrow\pm\infty$ . Then,  we are led to
${\delta\xi^{''}\phi/\delta\xi'}\ll$O(1) which is not acceptable.
 Finally, for
$\delta\xi^{''}z^2\gg$O(1)  we get from (11)
$z\sim{(\delta\xi^{'})^{-{2\over5}}}$  which is not consistent with
$z^2\delta\xi^{'}\gg1$ at $z\rightarrow\infty$. Thus, the case of $x_-$
is impossible. \par For $x={x_+}\sim{2\over{\delta\xi^{'}z}}$  we must
have $\vert{x/z}\vert\ll1$ . Then, equation (5) becomes
$$z^{'}\sim-{1\over2}\delta\xi^{'}z^3(1-{{8\over3}\delta\xi{''}
\over{z^6(\delta\xi^{'})^4}})\eqno(12)$$ 
 If $\delta\xi^{''}/z^6(\delta\xi^{'})^4\ll$O(1), we obtain
$$z^{2}\sim{1\over\delta\xi(\phi)-\delta\xi(\phi_s)}\eqno(13)$$ 
which is a
singular solution at some finite point $\phi_s$ . If
${\delta\xi^{''}}/{(\delta\xi^{'})^4}\gg{z^6}$, we can get from equation
(12) $z^4\sim{-{2\over3}/((\delta\xi^{'})^2+...)}$, which is
impossible as $z\rightarrow\infty$ . The only remaining case
${{\delta\xi^{''}}\over{(\delta\xi^{'})^4}}\sim{O(z^6)}$ can be studied
by putting $\delta\xi^{''}=b{z^6}{(\delta\xi^{'})^4}$.  Then, from (12)
we get $z^{-2}\sim (1-{8b\over3})(c+\delta\xi(\phi))$. Solving for
$\delta\xi^{'}$  we also get
$(\delta\xi^{'})^{-2}\sim const.+{b\over{(1-{8b\over3})^3
(c+\delta\xi)}}$ 
, which is equivalent to
${b{(\delta\xi^{'}z^2)^2}\over{({1-{8b\over3}})}}\sim1$ . This is
however, for any non-zero $b$, contradictory to our assumptions. \par
Thus, assuming that $\xi(\phi)$  is a smooth function that grows as fast
as $\phi^2$  at both infinities $\phi\rightarrow\pm\infty$, we have found
 only one
singular solution. Near the singular point $\phi_s$ 
$(\xi^{'}(\phi_s)\not=0)$ it behaves as

$$z\sim{(\delta\xi(\phi)-\delta\xi(\phi_s))^{-{1\over2}}}\eqno(14a)$$
$$x\sim{2\over{{\delta\xi^{'}(\phi)}z}}\eqno(14b)$$ 
  In terms of the
time,(14) becomes
$$\phi\sim{\phi_s+{{t^2}\over{\delta\xi^{'}(\phi_s)}}}\eqno(15a)$$
$$\omega\sim{\ln{t}}\eqno(15b)$$ where we have shifted the time variable
so that the singular point $\phi_s$ corresponds to t=0. Note that while
$z\rightarrow\infty$, $x$ goes to zero as long as
$\delta\xi^{'}(\phi_s)\not=0$ .

 The limit $x\rightarrow0$ of the system $(2),(3)$  deserves some
more attention. Indeed, as can be seen in equation (4), when
$x\rightarrow0$ for finite $\phi$ the only possibility for $z$ is to go
to zero too while $x\sim{\pm\sqrt{6}z}$ . Substituting the last
relation in equation (5), we obtain
${z^{'}/z}\sim{\mp\sqrt{3\over2}}$  or
$z\sim{\exp{\mp\sqrt{3\over2}\phi}}$  which does not go to zero
for finite $\phi$ in contradiction with our hypothesis. Inversely, when
$z\rightarrow0$ for finite $\phi$ the only possibility read off from
equation (4) is $x\sim{\pm\sqrt{6}z}$. Again, substituting that in
equation (5) we reach the same contradiction. The only conclusion we can
draw is that the points $x=0$ or $z=0$  cannot be reached analytically.
Therefore, the signs of $x$  and $z$  must be independently 
conserved. Then, the following statement is true for the singular
  solution we
found:
 Every connected piece of the solution will be characterized by a
fixed sign for each of $x$  and $z$ . We have assumed single-valuedness
of $z(\phi)$. If in a connected piece of the solution $z(\phi)$  were not
single-valued then at
 a non-singular point $\phi_1$ we would have
 $\frac{d\phi}{dz}=0$ at $z_1=z(\phi_1)$ . This is equivalent
to ${{xB}/{A{z_1^2}}}=0$  and it can happen only when either
$x\rightarrow0$ or $A\rightarrow\infty$ in contradiction to
$z_1<\infty$. 
  
Consider now a singular solution with the singularity
occurring at a point $\phi_s$. Each connected piece of the solution will
be characterized by a given sign of $x$  and $z$ . Take the piece with
$x>0$, $z>0$ . Then,
$$x=x_{-}\sim{2\over{z\delta\xi'}}\eqno(16)$$
 implies
that $\delta\xi^{'}>0$  and therefore, if $\delta>0$ ,$$\phi_s>\phi_0$$ 
where $\phi_0$  is the minimum of $\xi(\phi)$ . Because of
${z^{'}}\sim-{1\over2}{z^3}\delta\xi^{'}<0$  and the single-valuedness
of $z$ we come to the conclusion that these solutions cover at best the
full quarter-plane $z>0$, $\phi>\phi_0$  but they do not contain any
points of the $z>0$, $\phi<\phi_0$  region.
 We are now in a position to state the
following ``theorem" : 

{\it For $\delta>0$, any solution containing a point
 $\phi^{\ast}<\phi_0$ such that
$x(\phi^{\ast})>0$, $z(\phi^{\ast})>0$  must be a non-singular solution.} 
 
Since the solution is controlled by one first order differential
equation, namely (5), while (4) is just a soluble algebraic equation, it
depends only on $z$, $x$  at some initial point and we can always satisfy
the requirements of the theorem starting with values
$z^{\ast}>0$, $x^{\ast}>0$  at some initial point $\phi^{\ast}<\phi_0$ . 

\begin{figure}[htbp]
\epsfbox{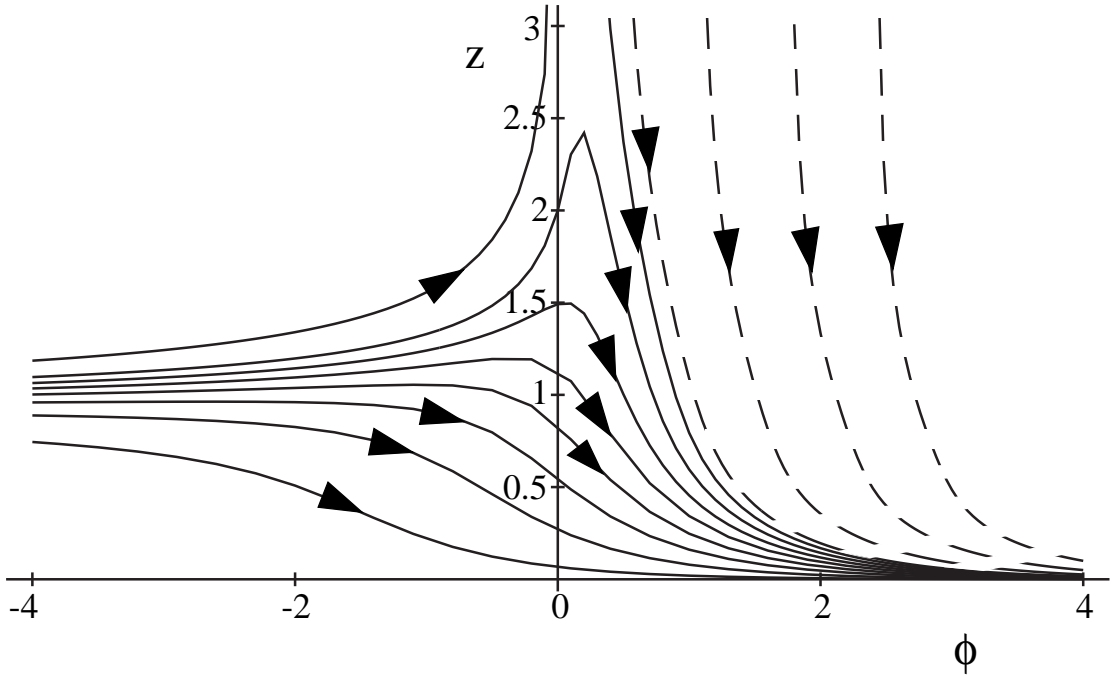}
\caption{The phase space diagram $z(\phi)$ for the case
$\xi(\phi)={1\over2}\phi^2$, $\delta=1$ and $z>0$, $x>0$. The dashed lines
correspond to singular solutions and the
 continuous lines to non-singular ones.
Arrows indicate the direction of the time.} 
\end{figure}

The existence of the non-singular
solution depends crucially on the fact that the singular solutions do
not cross over to the $\phi<\phi_0$  region. Notice that
 if we keep $x>0$, $z>0$
and chose $\delta<0$ 
 equation $(16)$ demands $\phi_s<\phi_0$ .
 Then, since $z^{'}<0$ still,the whole
upper plane is filled with singular solutions and there are no
singularity-free solution in this case.
 This behavior is in agreement with the
violation of the energy conditions required by singularity theorems
 $^{[5]}$ which cannot occur in the $\delta<0$ case since in this case
${\rho}+p$ and ${\rho}+3p$  are positive definite.

In order to make our analysis concrete we have also studied numerically the
case $\xi(\phi)=\phi^2/2$, $\delta=1$.
 The phase diagram for  $z=z(\phi)>0$ and
$x>0$ is presented in Fig. 1. As expected the singular solutions can occur
only for $\phi_s>\phi_0=0$ and points with $\phi<\phi_0=0$
 belong to non-singular
solutions. It can also be shown that in this specific case the non-singular
solutions interpolate between a De-Sitter space universe at the remote past
($t\rightarrow-\infty$) and a slowly expanding one at the future 
($t\rightarrow-\infty$).

The general behavior of the non-singular solutions at late
times $(t\rightarrow\infty)$ , assuming an expanding universe $(z>0)$ ,
is that of slow expansion, while the behavior at early
times$(t\rightarrow{-\infty})$
  depends crucially  on the particular choice of
$\xi(\phi)$. In the case of the loop corrected superstring effective
action $^{[1]}$  which has served as a motivation for the present investigation
\footnote{In the superstring effective action case [1], the modulus dependent
coefficient $\xi(\phi)$ of the
$R^2_{GB}$ term, satisfies the requirements of our proof, since
it has a unique minimum at $\phi=0$ and
 it grows as $\xi\sim e^{|\phi|}$ at
$\phi\rightarrow\pm\infty$.}
we obtain at early times a flat-space. In contrast, in the
case $\xi(\phi)={1\over2}{\phi^2}$, analyzed above ,  we have
obtained  De-Sitter space.
\smallskip

{\it Acknowledgments}
\smallskip 

One of us (K.T.) would like to acknowledge traveling
support from the EEC contract SCI-0394-L and the
University of Ioannina Research Committee.He would also like to 
thank the CERN Theory Division for its hospitality.
\bigskip
\vfil\eject
\centerline{\bf References}
\smallskip   

[1]     I. Antoniadis, J. Rizos and K. Tamvakis,
preprint hep-th|9305025, CPTH-A239.0593, IOA-291
(May 1993) , to appear in Nucl. Phys. {\bf B}.

[2]     A.A. Starobinsky,{\it Phys. Lett.}{\bf B91} (1980)99.

[3]     V. Mukhanov and R.Brandenberger,{\it Phys.Rev.Lett.}{\bf 68}
(1992)1969;\\R.Brandenberger,V.Mukhanov and A. Sornborger,
{\it Phys.Rev.}{\bf D48}(1993)1629. 
 
[4]     I. Antoniadis, E.
Gava and K.S. Narain,{\it Phys.Lett.}{\bf B283}(1992)209;\\
{\it Nucl..Phys.}{\bf B393}(1992)93. 

[5]     R. Penrose, {\it Phys.Rev.Lett.}{\bf 14}(1956)57;
S.Hawking, {\it Proc.R.Soc.London}{\bf A300}\\(1967) 182;
S.Hawking and R. Penrose, {\it Proc.R.Soc.London}{\bf A 314}
(1970)529.
\vfil\eject
 \end{document}